\documentstyle[aps,multicol,epsf,epsfig]{revtex}

\begin{document}

\begin{multicols}{2}

{\bf Comment on ``Small-world networks: Evidence for a crossover picture''}

In a recent letter, Barth\'el\'emy and Nunes Amaral 
\cite{bart} examine the crossover behaviour of networks known
as ``small-world''.
They claim that, for an initial network with $n$ vertices and
$z$ links per vertex, each link being rewired according
to the procedure of \cite{watts_strogatz} with a probability $p$,
the average distance $\ell$ between two vertices scales as
\begin{equation}
\ell(n,p) \sim n^* F \left( \frac{n}{n^*} \right)
\label{ansatz}
\end{equation}
where $F(u \ll 1) \sim u$, and $F( u \gg 1) \sim \ln u$, and $n^*$ 
$n^* \sim p^{-\tau}$ with $\tau = 2/3$ as $p$ goes to zero.

Other quantities can be of interest in small-world networks, and
will be discussed in \cite{inpreparation}. 
In this comment however, we concentrate like \cite{bart} on
$\ell$ and we show, using analytical arguments and numerical
simulations with larger values of  $n$, that:
(i) the proposed scaling form 
$\ell(n,p) \sim n^* F ( n/n^* )$ seems to be valid, BUT
(ii) the value of $\tau$ {\it cannot} be lower than $1$, and therefore
the value found in \cite{bart} is clearly wrong.

The naive argument developed in \cite{bart} uses the mean number of
rewired links, $N_r = p n z/2$. According to \cite{bart}, one could
expect that the crossover happens for $N_r = O(1)$, which gives
$\tau = 1$ \cite{note}. However they find $\tau = 2/3$. Let us suppose that
$\tau < 1$. Then, if we take $\alpha$ such that
$\tau < \alpha < 1$, according to eq (\ref{ansatz}), we obtain that
\begin{equation}
\ell(n,n^{-1/\alpha}) \sim n^{\tau/\alpha} F( n^{1-\tau/\alpha} )
\sim n^{\tau/\alpha} \ln (  n^{1-\tau/\alpha} )     
\end{equation}
since $\tau/\alpha < 1$ and  $n^{1-\tau/\alpha} \gg 1$ for large
$n$. However, the mean number of rewired links in this case is
$N_r = n^{1-1/\alpha} z/2 $, which goes to zero for large $n$. 
The immediate conclusion is that a change in the behaviour of
$\ell$ (from $\ell \sim n$ to $\ell \sim n^{\tau/\alpha} \ln (n)$) could
occur by the rewiring of a vanishing {\it number} of links!
This is a physical nonsense, showing that, if
$n^* \sim p^{-\tau}$, $\tau$ cannot be lower than $1$.

We know present our numerical simulations.
The value of $n^* (p)$ is obtained by studying, at fixed $p$ (we
take $p=2^k/2^{20}$, $k=0,\cdots 20$), the crossover
between $\ell \sim n$ at small $n$ to $\ell \sim \ln (n)$ at large $n$ 
\cite{bart}.
For small values of $p$, it is difficult to reach large enough values
of $n$ to accurately determine $n^*$, and we think that the 
underestimation of $n^*$ given by \cite{bart} comes from this problem.
We here simulate networks with $z=4,\ 6,\ 10$ up to sizes $n=11000$, and
find that $n^*$ behaves like $1/p$ for small $p$ (Inset of
fig. (1)). 
We moreover show the collapse of the curves $\ell/n^*$ versus
$n/n^*$ in figure (1), for $z=4$ and $z=10$: note that we obtain the collapse
over a much wider rabge than \cite{bart}. 

Besides, we present results for another quantity: at fixed $n$ 
we evaluate $\ell(n,p)$ and look for the value
$p_{1/2} (n)$ of $p$ such that 
$\ell(n,p_{1/2}(n)) = \ell(n,0) /2$.
This value of $p$ corresponds to the rapid drop in the plot of
$\ell$ versus $p$ at fixed $N$, and can therefore also be
considered as a crossover value.
If we note $u^*$ the number such that $F(u^*) = u^*/2$, then we obtain, since
$\ell(n,0) \sim n$, that $n^*(p_{1/2}(n)) = n/ u^*$.
If $n^* (p) \sim p^{-\tau}$, this implies
$p_{1/2}(n) \sim n^{-1/\tau}$. We show in fig (2) that 
$p_{1/2}(n) \sim 1/n$ (and clearly not $\sim n^{-3/2}$ like the results
of \cite{bart} would imply), meaning that $\tau =1$:
a finite number of rewired
links already has a strong influence  on $\ell$. Again 
the $\tau =2/3$ result
of \cite{bart} is clearly ruled out.

It is a pleasure to thank G. Biroli, R. Monasson and M. Weigt
for discussions.

A. Barrat

Laboratoire de Physique Th\'eorique \cite{lab},
Universit\'e de Paris-Sud,
91405 Orsay cedex,
France

\narrowtext
\begin{figure}
\centerline{ 
\epsfig{figure=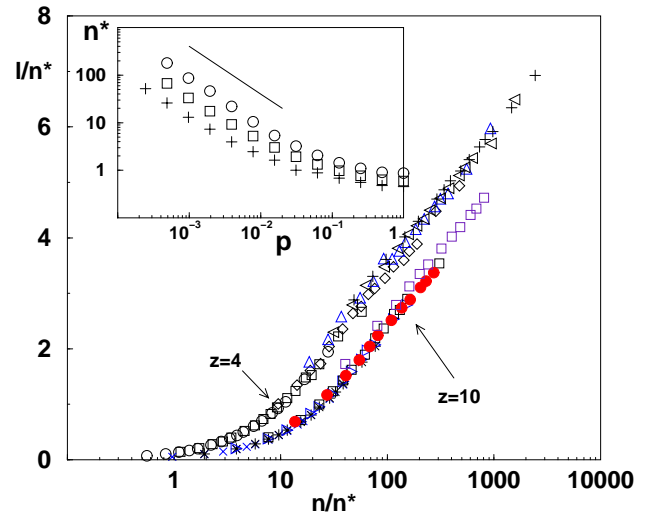,width=7cm,angle=-90}
       }
\caption{Data collapse of $\ell(n,p)/n^*$ versus
            $n/n^*$ for $z=4$ and $z=10$,
            for various values of $p$ and $n$ from $100$ to $5000$.
            Inset: $n^*$ versus $p$ for $z=4$ (circles), $6$ (squares),
	    $10$ (crosses); the straight line has slope $-1$.}
\label{fig:1}
\end{figure}

\begin{figure} 
\centerline{  
      \epsfig{figure=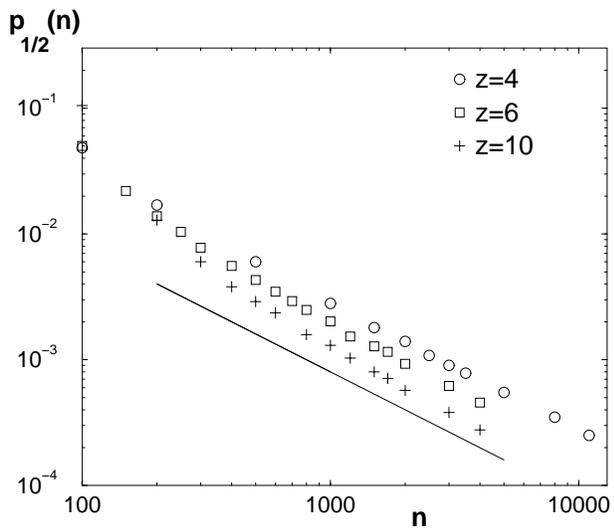,width=7cm,angle=-90}
       }
\caption{$p_{1/2}(n)$ such that 
         $\ell(n,p_{1/2}(n)) = \ell(n,0)/2$, for $z=4,\ 6,\ 10$, and
         values of $n$ ranging from $100$ to $11000$; the straight line 
	 has slope $-1$.}
\label{fig:2}
\end{figure}

\end{multicols}
\end{document}